\documentstyle[12pt,aasms4]{article}

\begin{document}

\noindent
{\it Dissertation Summary}

\begin{center}


\title{\large \bf The Star Formation Rate density of the Universe at $z$=0.24 and $z$=0.4 from H$\alpha$}

\end{center}


\author{Sergio Pascual}

\affil{Universidad Complutense de Madrid}

\begingroup

\parindent=1cm


\begin{center}

Electronic mail: spr@astrax.fis.ucm.es

Thesis work conducted at: Universidad Complutense de Madrid

Ph.D. Thesis directed by:  Jes\'us Gallego \&{} Jaime Zamorano;  ~Ph.D. Degree awarded: 2004-07-09


{\it Received \underline{\hskip 5cm}}

\end{center}

\endgroup


\keywords{Galaxies: distances and redshifts -- Galaxies: evolution --
Galaxies: luminosity function, mass function}


Knowledge of both the global star formation
history of the Universe and the nature of individual star-forming
galaxies at different look-back times are essential to our
understanding of galaxy formation and evolution. Deep redshift
surveys suggest star-formation activity increases by a order of magnitude from $z$=0 to $z\simeq1$.
To test directly whether substantial evolution in the
star-formation activity has occurred we need to measure the star formation rate (SFR)
density and the properties of the corresponding star-forming galaxy
populations at different redshifts using similar techniques.

The main goal of this thesis work\footnote{The original text is
written in Spanish and is available at
\mbox{http://t-rex.fis.ucm.es/Publications/Repository/tesis\_spr.ps.gz}} is to extend the survey
of emission-line galaxies of the {\it Universidad Complutense de Madrid}
(UCM survey\footnote{http://t-rex.fis.ucm.es/ucm\_survey}) to higher redshifts.

The H$\alpha$ line (6563\AA{}) luminosity, related to the number of massive stars, is
a direct measurement of the current star formation rate. 
To compare directly with the H$\alpha$-selected sample of the UCM survey, the selection of candidates
is made through narrow-band filters. The study of the colors of different objects (stars and galaxies)
in the system of narrow- and broad band filters shows strong increase of color as different emission lines
enter the narrow band filter. It is thus possible to make a feasible selection of emission line 
galaxies using narrow band filters. The relations between 
the broad-band magnitude and narrow-broad color with the line flux and the equivalent width of 
the emission line are developed. These relations are used to obtain the luminosity of the emission line, assuming an average obscuration and a mean value for
the doublet [\ion{N}{2}]$\lambda\lambda$6548,6584.

The night-sky air-glow spectrum has substantial gaps around 8200\AA{}
and 9200\AA, corresponding to H$\alpha$ redshifted respectively to $z\sim$0.24 and $z\sim$0.4. 
We have used two sets of filters. For the Wide Field Camera in Isaac Newton telescope at La Palma observatory, two narrow band filters (50{\AA} width) and centered at those wavelength were designed.
We also made use of a filter centered at 8200\AA{}, 150\AA{} width, with CAFOS,
in the 2.2m telescope at Calar Alto observatory.
Two samples of 
candidates at those redshifts have been built. The analysis of the objects is centered in the distribution
of equivalent width and luminosities, when compared with the UCM sample.
A important aspect of a survey with narrow-band filters  is that it is 
open to contaminant galaxies with 
other emission lines (mainly [\ion{O}{3}]$\lambda\lambda$4959,5007 and 
[\ion{O}{2}]$\lambda$3727) 
at higher redshifts. The fraction of contaminants is estimated analytically and found negligible at the 
relatively high line fluxes of the objects.

The SFR of the two samples is calculated assuming that every candidate selected in both filters 
(after stars removal) is a H$\alpha$ emitter. With this conditions, the SFRs are 3 times the SFR 
of the local Universe at $z\sim$0.24 and 5 times at $z\sim$0.40.
A sub-sample selected with the 8200\AA{} filter has been analyzed using photometric redshifts. We have found that 
some of the candidates appear at redshifts not covered with emission-lines, being misclassified objects. 
The luminosity density is then 
over-estimated in a factor $\sim1.2$. 
This correction is applied to the sample selected with the 9200\AA{}
filter and the SFR recalculated. 
The evolution of the SFR with redshift is found to be will fitted with a power-law 
$\rho\propto(1+z)^{\beta}$ with $\beta=4.1\pm0.1$

Spectroscopic follow-up observations have been made for the same sub-sample. 
Of the objects with H$\alpha$ confirmed by photometric redshifts, 12 (60\%) are spectroscopically confirmed. 
Two objects show the line [\ion{O}{3}]$\lambda\lambda$4959,5007 in emission, 
corresponding to a redshift about $z\sim$0.6. 
The presence of these two objects, at the relatively bright line flux limit of the sub-sample, exhibit
a stronger evolution in the luminosity functions than supposed in the analytical approach.

The galaxies confirmed with H$\alpha$ in emission are
mainly starburst-like (9 objects) and only 3 objects are HII-like. Only one object 
is classified as a LINER.
The masses, from rotation curves, 
are about the order of magnitude of the galaxies of the local Universe.
These galaxies represent a population very similar to the local population of star-forming 
galaxies selected with H$\alpha$.





\end{document}